\documentclass[onecolumn]{emulateapj}
\shorttitle{Photodesorption of Ices II: H$_2$O and D$_2$O}
\shortauthors{\"Oberg et al.}
\renewcommand{\baselinestretch}{2}
\begin{document}
%\doublespacing

%% LaTeX will automatically break titles if they run longer than
%% one line. However, you may use \\ to force a line break if
%% you desire.

\title{Photodesorption of Ices II: H$_2$O and D$_2$O\\[8ex]}

%% Use \author, \affil, and the \and command to format
%% author and affiliation information.
%% Note that \email has replaced the old \authoremail command
%% from AASTeX v4.0. You can use \email to mark an email address
%% anywhere in the paper, not just in the front matter.
%% As in the title, use \\ to force line breaks.

\author{Karin I. \"Oberg and Harold Linnartz}
\affil{Raymond and Beverly Sackler Laboratory for Astrophysics, Leiden Observatory, Leiden University, P.O. Box 9513, NL 2300 RA Leiden, The Netherlands}
\email{oberg@strw.leidenuniv.nl}

\and

\author{Ruud Visser and Ewine F. van Dishoeck\altaffilmark{1}}
\affil{Leiden Observatory, Leiden University, P.O. Box 9513, NL 2300 RA Leiden, The Netherlands}

%% Notice that each of these authors has alternate affiliations, which
%% are identified by the \altaffilmark after each name.  Specify alternate
%% affiliation information with \altaffiltext, with one command per each
%% affiliation.

\altaffiltext{1}{Also at Max-Planck Institut f\"ur Extraterrestrische Physik (MPE), Giessenbachstraat 1, D 85748 Garching, Germany}

%% Mark off your abstract in the ``abstract'' environment. In the manuscript
%% style, abstract will output a Received/Accepted line after the
%% title and affiliation information. No date will appear since the author
%% does not have this information. The dates will be filled in by the
%% editorial office after submission.

\begin{abstract}
Gaseous H$_2$O has been detected in several cold astrophysical environments, where the observed abundances cannot be explained by thermal desorption of H$_2$O ice or by H$_2$O gas phase formation. These observations hence suggest an efficient non-thermal ice desorption mechanism. Here, we present experimentally determined UV photodesorption yields of H$_2$O and D$_2$O ice and deduce their photodesorption mechanism. The ice photodesorption is studied under ultra high vacuum conditions and at astrochemically relevant temperatures (18--100~K) using a hydrogen discharge lamp (7-10.5 eV), which simulates the interstellar UV field. The ice desorption during irradiation is monitored using reflection absorption infrared spectroscopy of the ice and simultaneous mass spectrometry of the desorbed species. The photodesorption yield per incident photon, $Y_{\rm pd}(T,x)$, is identical for H$_2$O and D$_2$O and its dependence on ice thickness and temperature is described empirically by $Y_{\rm pd}(T,x)=Y_{\rm pd}(T,x>8)(1-e^{-x/l(T)})$ where $x$ is the ice thickness in monolayers (ML) and $l(T)$ a temperature dependent ice diffusion parameter that varies between $\sim$1.3~ML at 30~K and 3.0~ML at 100~K. For thick ices the yield is linearly dependent on temperature due to increased diffusion of ice species such that $Y_{\rm pd}(T,x>8) = 10^{-3}\left(1.3+0.032\times T\right)$ UV photon$^{-1}$, with a 60\% uncertainty for the absolute yield. The increased diffusion also results in an increasing H$_2$O:OH desorption product ratio with temperature from 0.7:1.0 at 20~K  to 2.0:1.2 at 100~K. The yield does not depend on the substrate, the UV photon flux or the UV fluence. The yield is also independent on the initial ice structure since UV photons efficiently amorphize H$_2$O ice. The results are consistent with theoretical predictions of H$_2$O photodesorption at low temperatures and partly in agreement with a previous experimental study. Applying the experimentally determined yield to a Herbig Ae/Be star+disk model provides an estimate of the amount of gas phase H$_2$O that may be observed by e.g. \textit{Herschel} in an example astrophysical environment. The model shows that UV photodesorption of ices increases the H$_2$O content by orders of magnitude in the disk surface region compared to models where non-thermal desorption is ignored.
\end{abstract}

%% Keywords should appear after the \end{abstract} command. The uncommented
%% example has been keyed in ApJ style. See the instructions to authors
%% for the journal to which you are submitting your paper to determine
%% what keyword punctuation is appropriate.

\keywords{Astrochemistry, Molecular data, Molecular processes, Methods: laboratory, Ultraviolet: ISM, ISM: molecules, Circumstellar matter}

\section{Introduction}

H$_2$O, in solid or gaseous form, is one of the most common species in molecular clouds, typically only second to H$_2$ and sometimes to CO. This makes H$_2$O, together with CO, the dominant reservoir of oxygen during the critical stages of star formation \citep{vanDishoeck93}. H$_2$O is thus a key molecule in astrochemical models and its partitioning between the grain and gas phase therefore has a large impact on the possible chemical pathways, including the formation of complex organics \citep{Charnley92, vanDishoeck06b}.

In cold, quiescent clouds H$_2$O forms through hydrogenation of O (O$_2$ or O$_3$) on cold (sub)micron-sized silicate grain surfaces forming icy layers \citep{Tielens82,Leger85,Boogert04,Miyauchi08,Ioppolo08}. Other ices, like NH$_3$ and CH$_4$, probably form similarly, but observations show that H$_2$O is the main ice constituent in most lines of sight, with a typical abundance of $1\times 10^{-4}$ with respect to the number density of hydrogen nuclei. Gas phase H$_2$O formation is only efficient above 300 K \citep{Elitzur78,Elitzur78b,Charnley97}. At lower temperatures gas phase ion-molecule reactions maintain a low H$_2$O abundance around 10$^{-7}$ \citep{Bergin95}. Any higher abundances require either thermal desorption of the H$_2$O ice above $\sim$100 K or non-thermal desorption at lower temperatures \citep{Bergin95, Fraser01}.

Gas phase H$_2$O is observed from the ground only with great difficulty. Still both isotopic and normal H$_2$O have been detected in astrophysical environments from ground based telescopes \citep{Jacq88,Knacke91,Cernicharo90,Gensheimer96,vanderTak06}. The {\it Infrared Space Observatory (ISO)} detected warm H$_2$O gas unambiguously toward several low- and high-mass young stellar objects \citep{vanDishoeck96,Ceccarelli99, Nisini99,Boonman03}. {\it ISO} was followed by two other space based telescopes, the {\it Submillimeter Wave Astronomy Satellite (SWAS)} and {\it Odin}. In difference to {\it ISO}, {\it SWAS} and {\it Odin} are capable of detecting the fundamental ortho-H$_2$O 1$_{10}-1_{01}$ transition at 538.3 $\mu$m and hence probe cold H$_2$O gas \citep{Melnick00,Hjalmarson03}. Both telescopes have observed H$_2$O gas toward star forming regions and detected the expected enhancements near protostars and in outflows where thermal ice desorption or gas phase formation is possible \citep{Hjalmarson03,Franklin08}. Critical for the present study, H$_2$O gas has also been detected toward photon dominated regions \citep{Snell00,Wilson03} and is also more abundant toward diffuse than toward dense clouds. These two results point to an efficient ice photodesorption mechanism \citep{Melnick05}. The importance of photodesorption at the edges of clouds has more recently  been modeled by Hollenbach et al. (ApJ, in press). They find that the H$_2$O gas abundance is enhanced by orders of magnitude at A$_{\rm V}$=2--8 mag into the cloud when including photodesorption of H$_2$O ice in their model at a rate derived from the results by \citet{Westley95}. Circumstellar disks is a second region where the impact of photodesorption is expected to be be large. \citet{Willacy00} showed that a photodesorption yield of 10$^{-3}$ molecules per photon is enough to explain observed gas-phase CO abundances in the outer regions of flared disks. Employing a similar photodesorption yield for H$_2$O ice, the disk models of \citet{Dominik05} and \citet{Willacy07} both predict the existence of significant amounts of gas phase H$_2$O in a layer above the midplane region.

With the advent of the {\it Herschel Space Observatory}, cold and warm H$_2$O gas observations on scales of protostellar envelopes and disks will for the first time become possible \citep{vanKempen08}. In preparation for these and other observations, and to interpret data from {\it Odin} and {\it SWAS}, non-thermal processes need to be better understood. These non-thermal desorption processes include ion/electron sputtering, desorption due to the release of chemical energy and photodesorption. Of these, sputtering of ice by electrons and ions, has been investigated over a range of conditions during the past few decades \citep[e.g.][]{Brown78, Fama08} and the dependencies of the yield on e.g. ice temperature, projectile type and energy are rather well understood. In contrast, only a handful of laboratory studies exists on the efficiency of ice photodesorption \citep[e.g.][]{Westley95, Oberg07b, Oberg08b}.  

\citet{Westley95, Westley95b} determined the photodesorption rate of H$_2$O ice experimentally to be $3-8\times10^{-3}$ molecules per UV photon for a 500 nm thick H$_2$O ice. In their experiment the photodesorption rate depends on UV fluence as well as temperature. The photon fluence dependence, together with the observed gas phase H$_2$ and O$_2$ during irradiation, was taken as evidence that H$_2$O photodesorption at low temperatures mainly occurs through desorption of photoproducts rather than of H$_2$O itself. Several questions remain regarding the applicability of their study to astrophysical regions due to the uncertainty of the proposed mechanism. In addition, their dependence on photon fluence is not reproduced in recent CO and CO$_2$ photodesorption experiments and cannot easily be explained theoretically \citep{Andersson06, Oberg07b, Andersson08, Oberg08b}. 

In a different experiment \citet{Yabushita06} investigated H-atom photodesorption from H$_2$O ice during irradiation at 157 and 193 nm using time-of-flight mass spectrometry. They found that the temperature and hence the origin of the desorbed H atoms varies significantly between crystalline and amorphous ice at 100 K. This indicates that photodesorption depends on the ice morphology, which is in contrast to the findings of \citet{Westley95}.  Desorption of recombined D$_2$ during irradiation of D$_2$O ice at 12 K has also been found by \citep{Watanabe00}. Both experiments provide valuable input for models, but cannot directly be used to determine the total H$_2$O photodesorption rate.

UV irradiation of H$_2$O ice results in photochemistry products as well as photodesorption. A variety of photochemistry products are readily produced during irradiation of H$_2$O dominated ice mixtures \citep{Dhendecourt82, Allamandola88}. Pure H$_2$O ice photolysis results in the production of H$_2$O$_2$, OH radicals and HO$_2$ at 10 K \citep{Gerakines96,Westley95b}. After a fluence of $\sim5\times10^{18}$ UV photons cm$^{-2}$ \citet{Gerakines96} found that the final band area of the formed H$_2$O$_2$  was only $\sim$0.25\% compared to the H$_2$O band area and the OH band area was even smaller.  No study exists for higher temperatures, but as photodissociation fragments become more mobile, e.g. O$_2$ formation would be expected \citep{Westley95}.

Only a handful models of ice photodesorption exists in the literature. \citet{Andersson06,Andersson08} have investigated H$_2$O photochemistry and photodesorption theoretically using classical dynamics calculations. In the simulations they followed H$_2$O dissociation fragments, after the absorption of a UV photon, in the top six monolayers of both crystalline and amorphous H$_2$O ice at 10 K. For each ice they found that desorption of H$_2$O has a low probability (less than 0.5\% yield per absorbed UV photon) for both types of ice. The total H$_2$O photodesorption yield from the top six ice layers was calculated to be $\sim4\times10^{-4}$ molecules per incident UV photon.

In the present study we aim at determining experimentally the photodesorption yields of H$_2$O and D$_2$O and their dependencies on ice thickness, temperature, morphology, UV flux and fluence as well as irradiation time. We use these results as input for an astrophysical model of a typical circumstellar disk to estimate the impact of photodesorption and to predict the observable column densities of H$_2$O as relevant to e.g. upcoming {\it Herschel} programs.

\section{Experiments and data analysis}

\subsection{Experiments}

\begin{deluxetable*}{llccc}
%\tabletypesize{\scriptsize}
\tablecaption{Summary of H$_2$O and D$_2$O experiments \label{h2oexps}}
\tablewidth{0pt}
\tablehead{\colhead{Experiment} & \colhead{Composition} & \colhead{Temperature [K]} & \colhead{Thickness [ML]} & \colhead{UV Flux [$10^{13}$ cm$^{-2}$ s$^{-1}$]} 
%& &  &\colhead{K} & \colhead{ML} & \colhead{$10^{13}$ cm$^{-2}$ s$^{-1}$}
}
\startdata
1&H$_2$O&18&14&2.3\\
2&H$_2$O&18&15&3.5\\
3&H$_2$O&100&12&1.1\\
4&H$_2$O&100&13&5.0\\
5&H$_2$O&100&17&2.3\\
6&D$_2$O&18&10&2.3\\
7&D$_2$O&18&11&3.5\\
8&D$_2$O&18&17&2.3\\
9&D$_2$O&30&2.1&3.5\\
10&D$_2$O&30&3.2&3.5\\
11&D$_2$O&30&8.9&3.5\\
12&D$_2$O&30&11&3.5\\
13&D$_2$O&40&6&1.7\\
14&D$_2$O&40&14&1.1\\
15&D$_2$O&40&14&3.5\\
16&D$_2$O&60&9.6&1.1\\
17&D$_2$O&60&17&3.5\\
18&D$_2$O&100&1.5&1.9\\
19&D$_2$O&100&2.3&1.8\\
20&D$_2$O&100&5.1&2.3\\
21&D$_2$O&100&5.3&2.3\\
22&D$_2$O&100&6.8&1.8\\
23&D$_2$O&100&12&5.0\\
24&D$_2$O&100&13&1.1\\
25&D$_2$O&100&14&3.5\\
26&D$_2$O&100&16&2.3\\
27&D$_2$O&100&28&1.7\\
28&D$_2$O\tablenotemark{a}&30&14&2.3\\
29&H$_2^{18}$O&20&14&5.0\\
30&H$_2^{18}$O&30&14&5.0\\
31&H$_2^{18}$O&100&14&5.0\\
32&D$_2$O/H$_2$O\tablenotemark{b}&30&9.5/48&2.3\\
33&N$_2$/D$_2$O\tablenotemark{b}&18&20/14&2.3\\
\enddata
%% Text for table notes should follow after the \enddata but before
%% the \end{deluxetable}. Make sure there is at least one \tablenotemark
%% in the table for each \tablenotetext.
\tablenotetext{a}{Annealed at 100 K for 1 hour}
\tablenotetext{b}{Ice layers}
\end{deluxetable*}

The experimental set-up (CRYOPAD) is described in detail by \citet{Fuchs06} and \citet{Oberg07b}. The set-up allows simultaneous detection of molecules in the gas phase by quadrupole mass
spectrometry (QMS) and in the ice by reflection absorption infrared
spectroscopy (RAIRS) using a Fourier transform infrared (FTIR) spectrometer. The FTIR covers 1200 -- 4000 cm$^{-1}$ with a typical spectral resolution of 1 cm$^{-1}$. The experimental procedure to derive photodesorption yields is described extensively in \citet{Oberg08b}, where the photodesorption measurements of CO, N$_2$ and CO$_2$ ices are reported. Here the procedure is summarized and only modifications to the procedure are described in detail. 

In the experiments, H$_2$O and D$_2$O ices of 1.5--28 monolayers (ML) are grown under ultra high vacuum conditions ($P\sim10^{-9}$ mbar with the background pressure dominated by H$_2$) at 18 -- 100~K on a gold substrate that is mounted on the coldfinger of a He cryostat. The H$_2$O sample is prepared from de-ionized H$_2$O that is purified through at least three freeze-thaw cycles. The D$_2$O sample is measured to have a 90\% isotopic purity and is similarly freeze-thawed before use.

Within the experimental uncertainties, we find that there is no difference in the photodesorption rate of 9.5 ML D$_2$O ice deposited on top of 48 ML H$_2$O ice compared with 8.9 ML D$_2$O ice deposited directly onto the gold substrate. Since the nature of the substrate has no influence on the photodesorption, all other experiments are carried out with H$_2$O or D$_2$O ices deposited directly on the gold substrate. 

The ice films are irradiated at normal or 45$^{\circ}$ incidence with UV light from a broadband hydrogen microwave discharge lamp, which peaks around Ly $\alpha$ at 121 nm and covers 115--170 nm or 7--10.5 eV \citep{Munozcaro03}. The lamp UV photon flux is varied between 1.1 and 5.0 $\times 10^{13}$ photons cm$^{-2}$ s$^{-1}$ in the different experiments. The lamp flux is measured by a NIST calibrated silicon diode and actiometry as described in \citet{Oberg08b}. 

Table \ref{h2oexps} summarizes the experiments in this study. The majority of the photodesorption experiments is carried out with D$_2$O ice. These experiments are complimented with H$_2$O and H$_2^{18}$O experiments to test isotope effects and to ensure the validity of mass spectrometric detections of OH (OD) and H$_2$O (D$_2$O) fragments. Layered experiments with H$_2$O and D$_2$O at 30 and 100 K and with N$_2$ on top of D$_2$O at 18 K are performed to check for substrate effects and to determine the ice loss behaviour when desorption is hindered.

\subsection{Data Analysis}

The UV induced ice loss rate during each H$_2$O and D$_2$O experiment is determined by RAIRS of the ice as a function of UV fluence. The intensity of the RAIRS profile is linearly correlated with the ice layer thickness up to $\sim$20 ML, but the RAIRS profile can be used up to 50 ML for analysis as long as the non-linear growth above 20 ML is taken into account \citep{Oberg08b}. One monolayer is generally taken to consist of $\sim$10$^{15}$ molecules cm$^{-2}$ and the loss yield, in molecules photon$^{-1}$, of the original ice is subsequently derived from the intensity loss in the RAIR spectra as a function of fluence.  

The determined ice loss yield is not necessarily the photodesorption yield. H$_2$O has only dissociative transitions in the wavelength region of the lamp. Hence, the UV irradiation induces photodesorption as well as photochemistry \citep{Gerakines96, Westley95}. UV irradiation may also induce structural changes in the ice that modify the infrared spectral features. These bulk processes, photolysis and rearrangement, are separated from the photodesorption by exploiting the different kinetic order behavior of bulk processes and surface desorption, i.e. first versus zeroth order processes. This method is described in detail in \"Oberg et al. (2008).

Using RAIRS to determine the desorption of molecules depends on a reliable conversion between the ice infrared absorbance and the amount of ice molecules. Due to the fact that all ice measurements are done using RAIRS, the ice thickness cannot be estimated from previously determined ice transmission band strengths. In \citet{Oberg08b}, the CO and CO$_2$ appropriate RAIRS band strengths are reported. The H$_2$O and D$_2$O band strengths are estimated by assuming that the relative band strengths of CO, CO$_2$ and H$_2$O ice are the same in transmission and reflection-absorption spectra. This is found to be accurate within a factor of two by \citet{Ioppolo08}. The thickness uncertainty is then $\sim$50\%. For the conversion between H$_2$O and CO and CO$_2$ band strengths, the measured band strengths of \citet{Hudgins93} are used after modification as suggested by \citet{Boogert97}. The relative band strengths of H$_2$O and D$_2$O were measured by \citet{Venyaminov97}. This results in H$_2$O and D$_2$O stretching band strengths of 0.95 and 0.68 cm$^{-1}$ ML$^{-1}$, respectively, for our set-up. These band strengths are converted to cm molecule$^{-1}$ assuming a monolayer density of 10$^{15}$ molecules ML$^{-1}$ cm$^{-2}$.

Kinetic modeling of the integrated RAIRS profiles as a function of UV fluence and the determined band strengths together provide a total ice photodesorption yield. The simultaneous mass spectrometry of gas phase molecules during irradiation reveals the nature of the desorbed species, i.e. what proportion of H$_2$O ice desorbs as H$_2$O molecules versus photo-produced radicals and molecules. This is limited by the fact that less volatile molecules adsorb onto the heating shield and other semi-cold surfaces inside the experiment before reaching the mass spectrometer. Hence, the relative abundance of species with very different cryopumping rates, like H$_2$ and H$_2$O, cannot be derived. It is however possible to estimate the ratio of the predicted main desorption species: H$_2$O and OH.

The main sources of uncertainty in these experiments are the photon flux at the sample surface and ice thickness calibrations of $\sim$30\% and 50\%, respectively. In addition, from repeated experiments, the H$_2$O experimental results vary with 20\%. The total uncertainty is hence $\sim$60\% for the total photodesorption rate. The relative desorption yields of different desorption products is more uncertain due to the additional assumptions that go into their derivation, i.e. that the products have similar pumping rates and QMS detection efficiencies. We estimate that the relative desorption yields thus have $\sim$30\% uncertainty in addition to the uncertainty of the total photodesorption yield.

\section{Results}

\subsection{Photodesorption process and products}

\begin{figure}
%\epsscale{.80}
\plotone{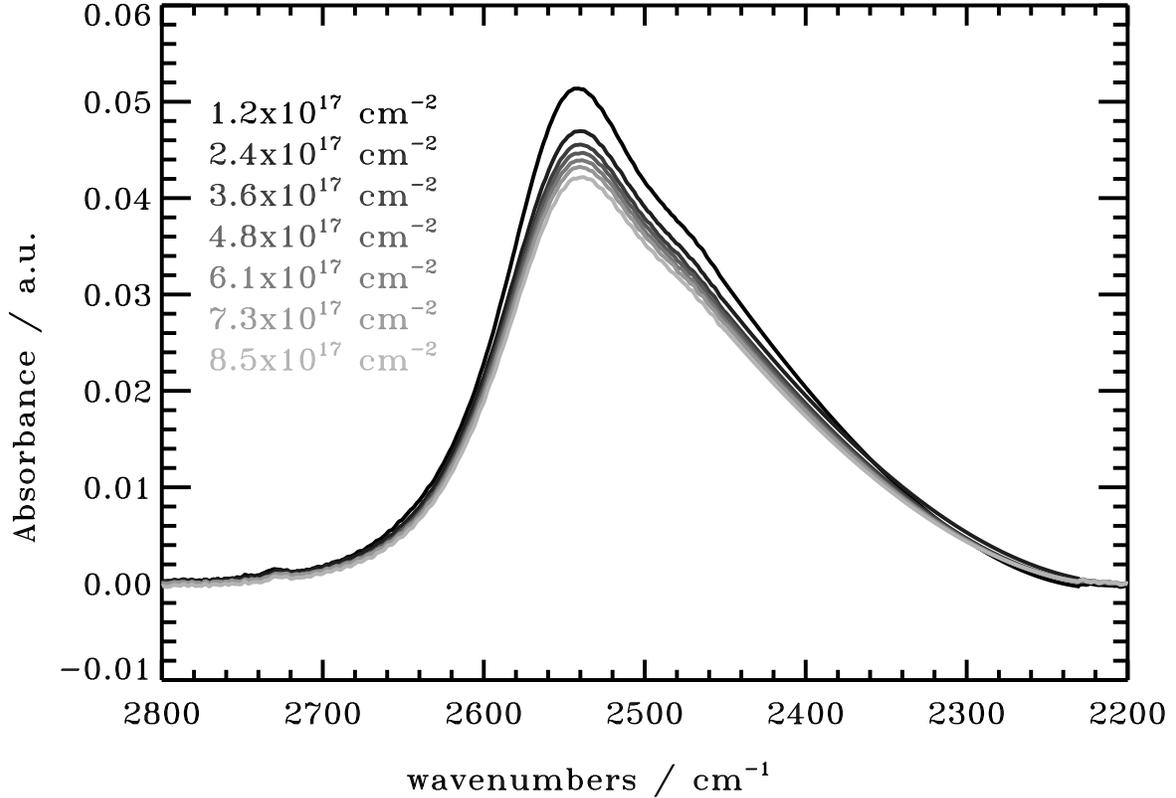}
\caption{Spectra of the D$_2$O stretching mode at 40~K as a function of UV fluence. The decreasing ice loss yield with fluence is due to the fact that ice is lost through a combination of bulk photolysis and photodesorption. The small bump at 2730 cm$^{-1}$ is probably caused by the free OD stretch.
\label{sp40k}\\[2ex]}
\end{figure}

\begin{figure}
\plotone{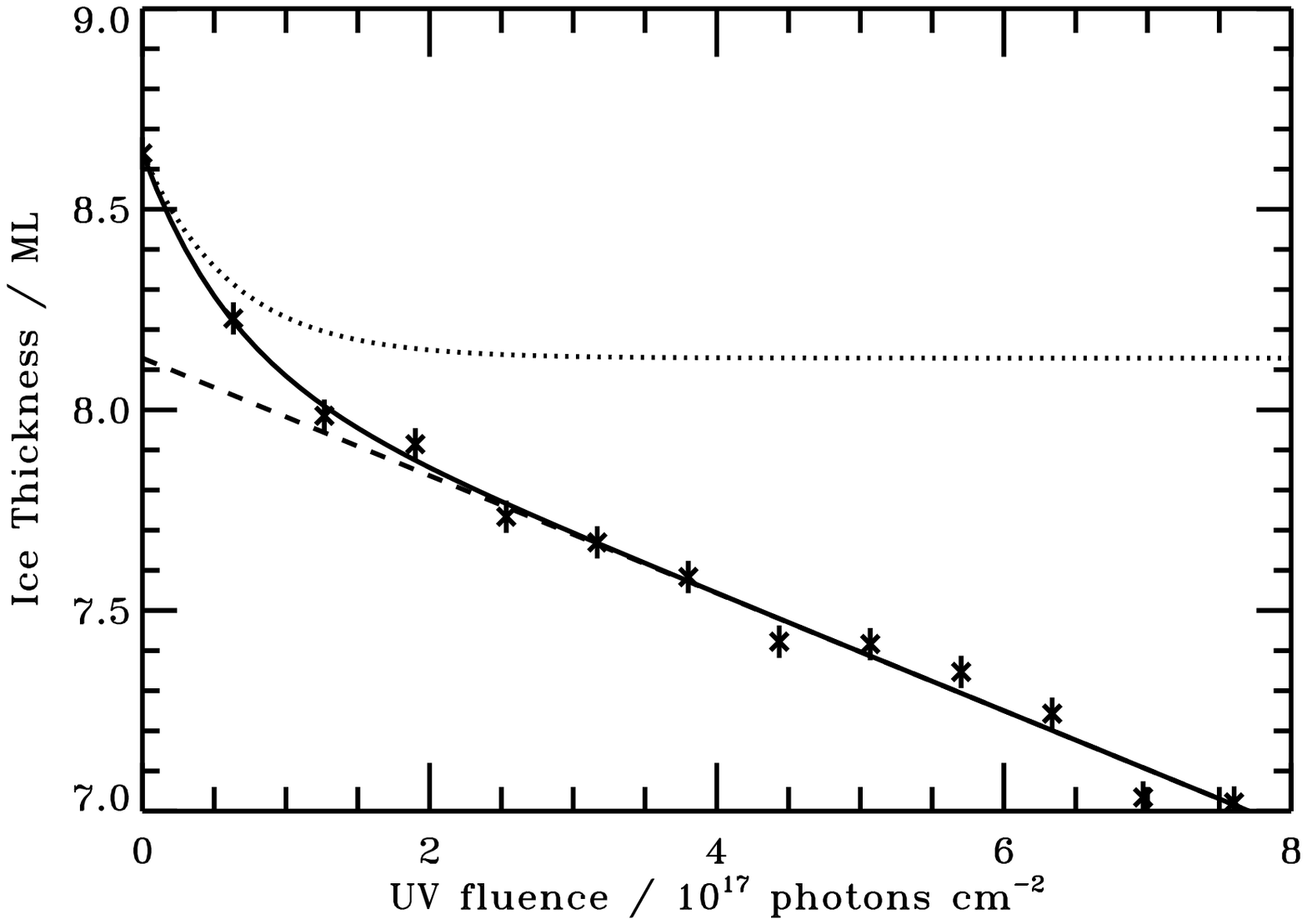}
\caption{The D$_2$O ice layer thickness as a function of UV fluence together with the fit (solid line), which is split up in ice loss due to bulk photolysis (dotted line) and photodesorption (dashed line). The fitted bulk photolysis contribution is offset for visibility.
\label{fit40k}\\[2ex]}
\end{figure}

Figure \ref{sp40k} shows the RAIR spectrum of the D$_2$O stretching band at 40 K as a function of UV fluence. At 40 K the UV photons simultaneously induce dissociation of bulk D$_2$O ice and desorption of surface molecules. This is clearly seen in Fig. \ref{fit40k}, where the ice thickness is plotted versus UV fluence. The ice loss is modelled by a combination of an exponential function and a linear function, where the exponential function describes the photolysis of bulk D$_2$O as a first order process similarly to \citet{Cottin03}.  The linear part of the ice loss is interpreted as photodesorption of surface molecules, which should be a zeroth-order process \citep{Oberg07b}. To test that this model holds, a D$_2$O ice is irradiated at 18~K when covered by a thick N$_2$ ice, which hinders desorption. The ice loss curve is then very well fitted ($\chi^2$=1.8 for 13 data points) by the exponential function derived from a bare 18 K ice. This shows that the observed exponential decay is indeed due to bulk photolysis with a possible contribution from ice rearrangement. As discussed below the ice structure is affected by the UV irradiation, but this rearrangement seems complete within an UV fluence of $1\times10^{17}$ photons cm$^{-2}$. It may hence contribute somewhat to the exponential decay of the RAIRS profile, but is completely filtered out from the photodesorption rate determination.

There are no other species than H$_2$O (D$_2$O) visible in the RAIR spectra at any temperature. OH (OD) formation cannot be excluded, however, due to the spectral overlap of OH (OD) and H$_2$O (D$_2$O) transitions. Despite this overlap, photolysis of H$_2$O into OH will result in a measured decrease of the H$_2$O stretching band area. This is both because of the expected lower band strength of OH compared to H$_2$O and because the remaining H$_2$O stretching band strength decreases when the H$_2$O network is disturbed by other molecules or fragments \citep{Oberg07,Bouwman07}. The lack of H$_2$O$_2$ formation is in contrast to e.g. \citet{Gerakines96}. This is not a contradictory result however, since less than 0.1\% of the H$_2$O ice is expected to be converted after a similar fluence, which is close to the detection limit here for the strongest H$_2$O$_2$ band at 2850 cm$^{-1}$ \citep{Giguere59}. 

\begin{figure}
\plotone{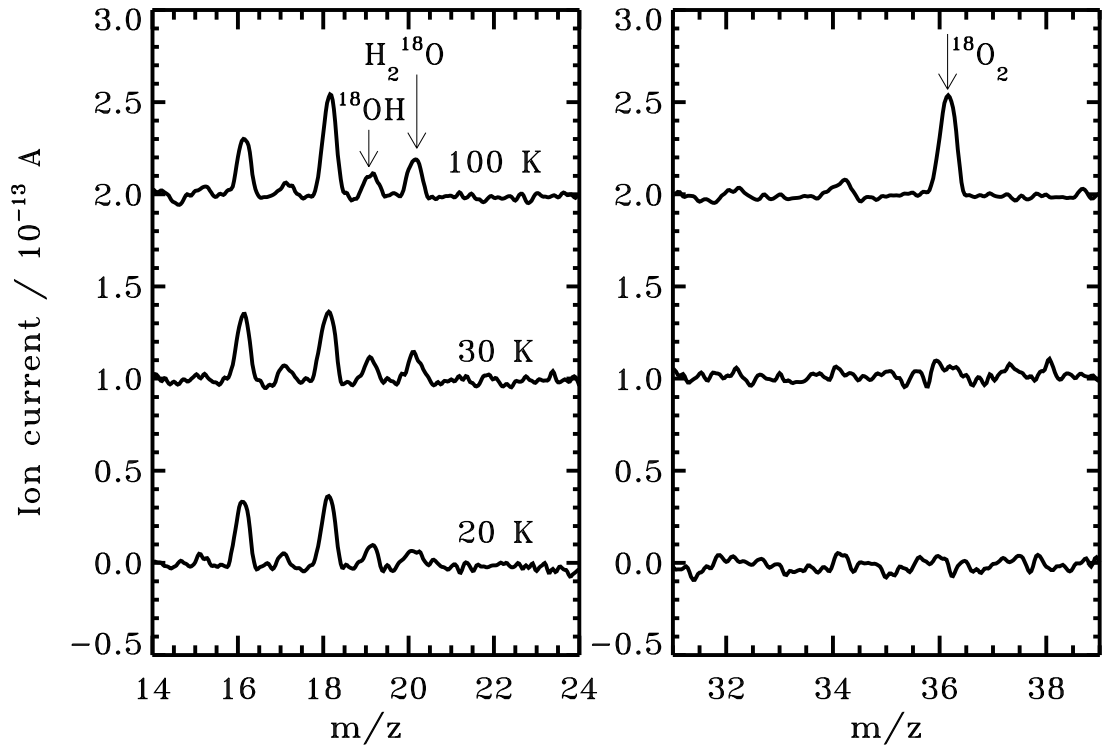}
\caption{The mass spectra recorded during irradiation of a 14 ML thick H$_2 ^{18}$O ice at 20, 30 and 100~K. In addition to photodesorbed ice product signals, background signals from H$_2$O, OH, O and O$_2$ are also seen.
\label{ms_sp}\\[2ex]}
\end{figure}

During irradiation H$_2$ (D$_2$) is always detected by the mass spectrometer. At the highest fluxes OH (OD) and H$_2$O (D$_2$O) are detected as well. A 14 ML H$_2 ^{18}$O ice is irradiated at 20, 30 and 100 K while acquiring mass spectra (Fig. \ref{ms_sp}) to quantify the relative desorption amounts of OH and H$_2$O without any overlap with background H$_2$O. The figure shows that the desorbing fraction of OH and H$_2$O changes somewhat with temperature from 1.0:0.7 at 20 K to 1.2:1.4 at 30 K to 1.2:2.0 at 100 K. No other species are observed at 20 and 30 K. In contrast, at 100 K, O$_2$ is photodesorbed as well. It is important to note that O$_2$ and H$_2$O have very different cryopumping rates and hence their relative mass spectromeric signals are not representative of their relative desorption rates. The upper limit of O$_2$ desorption is estimated from the fact that a factor of 1.9 more OH and H$_2$O is detected at 100 K compared to at 20 K, while the total photodesorption rate from RAIRS increases by a factor of 2.4. Hence at most one fifth of the H$_2$O photodesorbs as O$_2$ at high temperatures. Desorption of H$_2$O$_2$ cannot be excluded, even though it is not detected, since it is notoriously difficult to detect with a QMS. No build-up of H$_2$O$_2$ is observed in the ice, which makes it unlikely that a large part of the ice is desorbed in the form of H$_2$O$_2$.  Disregarding the small amount of H$_2$O that does not desorb as either OH or H$_2$O, the H$_2$O yield relative to the total photodesorption yield is fitted linearly as a function of temperature for the 14 ML thick ice . This empirical fit yields an expression for the H$_2$O yield, $Y_{\rm pd,H_2O}$, as a function of  the total photodesorption yield, $Y_{\rm pd}$:

\begin{eqnarray}
Y_{\rm pd,H_2O}=f_{\rm H_2O}\times Y_{\rm pd} \\
f_{\rm H_2O} =  (0.42\pm0.07)+(0.002\pm0.001)\times T\\
f_{\rm H_2O} + f_{\rm OH}\sim 1
\label{frac}
\end{eqnarray}

\noindent where $f_{\rm x}$ is the fraction of the total photodesorption that occurs through species $x$. The relative yields are probably somewhat thickness dependent \citep{Andersson06}, but due to experimental constraints it is not possible to probe the relative yields for thinner ices.

\subsection{Yield dependencies on temperature, fluence, ice thickness, flux and isotope}

\subsubsection{Temperature and photon fluence} 

\begin{figure}
\plotone{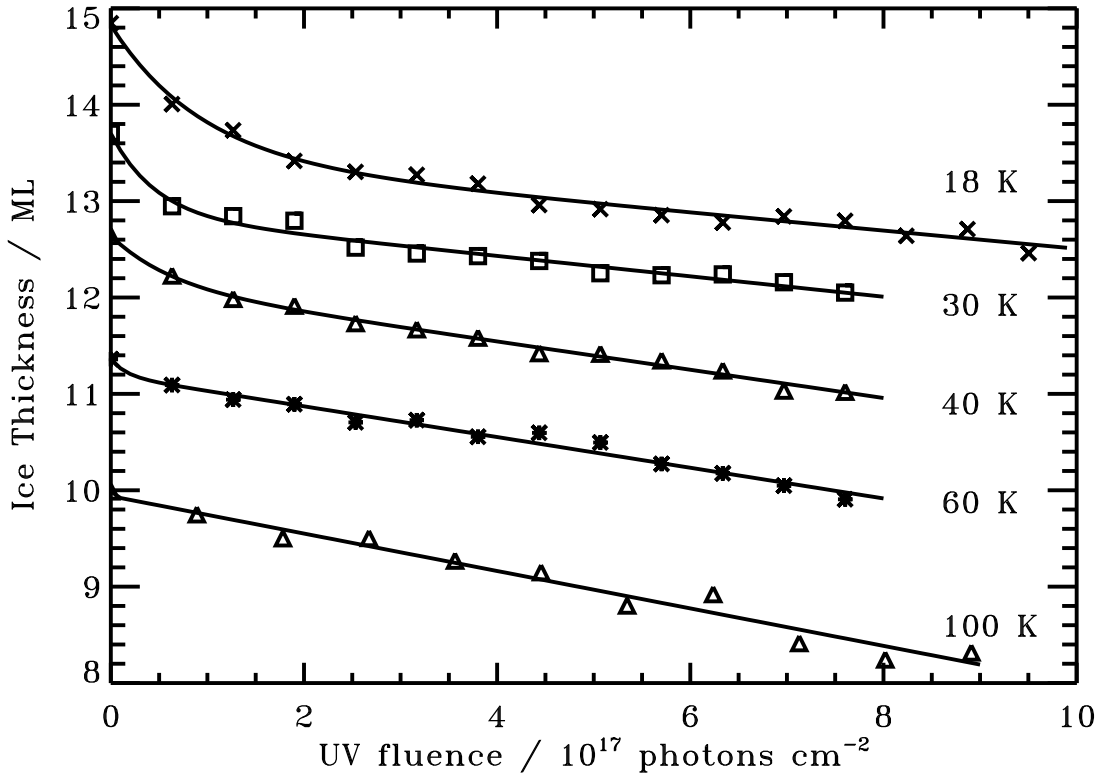}
\caption{Ice thickness (in ML) versus photon fluence (in 10$^{17}$ photons cm$^{-2}$) for $\sim$10~ML D$_2$O ices at different temperatures, displaying the temperature dependent degree of ice bulk photolysis versus ice photodesorption. The curves are offset for visibility. At lower temperatures the bulk photolysis dominates the ice loss, while it is not visible above 60 K. Simultaneously the photodesorption rate, the slope of the linear part of the ice loss, increases with temperature.
\label{temp}\\[2ex]}
\end{figure}

\begin{figure*}
\plotone{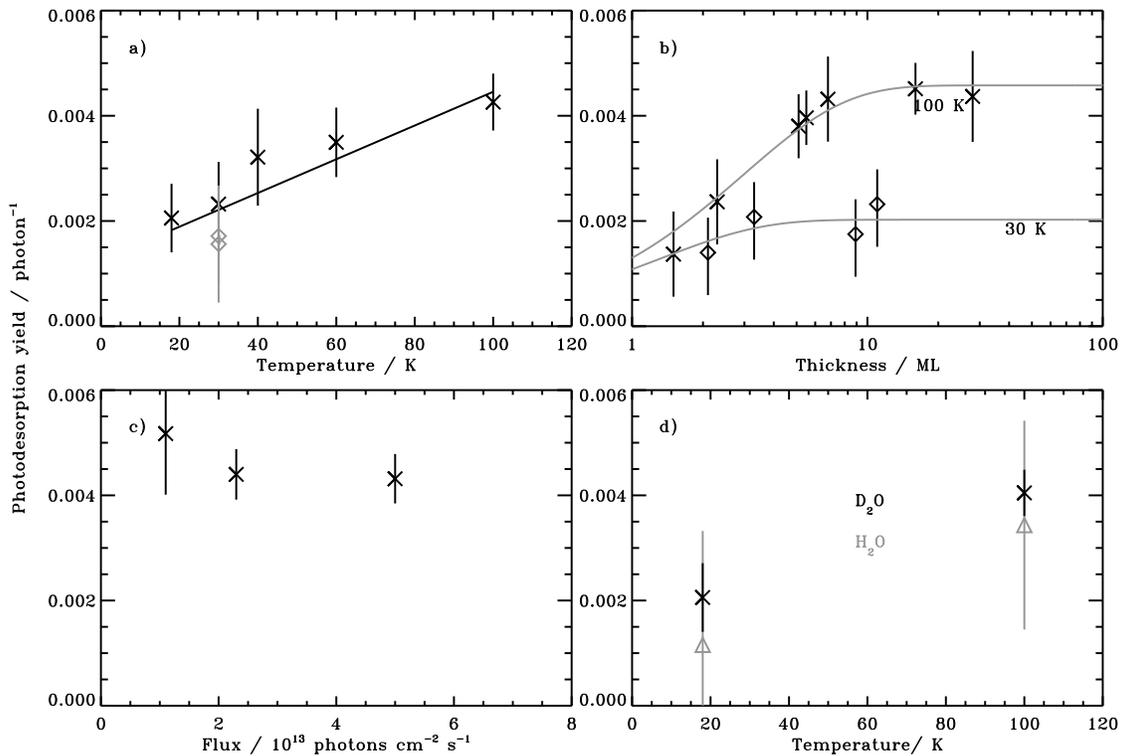}
\caption{The D$_2$O photodesorption yield as a function of temperature (a), ice thickness (b), lamp flux (c) and isotope (d). In the temperature plot (a) the ices are $\sim$10 ML thick. The two grey diamonds mark the desorption rate from a D$_2$O:H$_2$O layered ice and an annealed ice. In panel (b) the thickness dependence is plotted and fitted for ices at 30 and 100 K. In panel (c) the ices are 12--16 ML thick and irradiated at 100 K. The experiments marked with grey triangles in panel (d) are carried out with H$_2$O instead of D$_2$O. 
\label{depend}\\[2ex]}
\end{figure*}

Figure \ref{temp} shows the combined bulk photolysis and photodesorption for $\sim$10 ML thick D$_2$O ices between 18 and 100 K. The ice loss is dominated by bulk photolysis at low temperatures and by photodesorption at higher temperatures. This conclusion follows from the observed facts that 1) the degree of steady state photolysis decreases with temperature and 2) that the photodesorption yield per incident photon, as measured by the slope of the linear part of the fit, increases with temperature. This increase in photodesorption yield with temperature is shown explicitly in Fig. \ref{depend}a. Between 18 and 100 K the dependence of the photodesorption yield on temperature for thick ices ($>$8 ML) is empirically fitted with a linear function: 

\begin{equation}
Y_{\rm pd}(T,x>8) = 10^{-3} \left(1.3\left(\pm 0.4\right) + 0.032\left(\pm 0.008\right)\times T\right)
\label{tempeq}
\end{equation}

\noindent where $T$ is the temperature in K and $x$ the ice thickness in ML. The uncertainties are the model fit errors -- the total uncertainty of the yield is 60\% as stated above.

\begin{figure}
\plotone{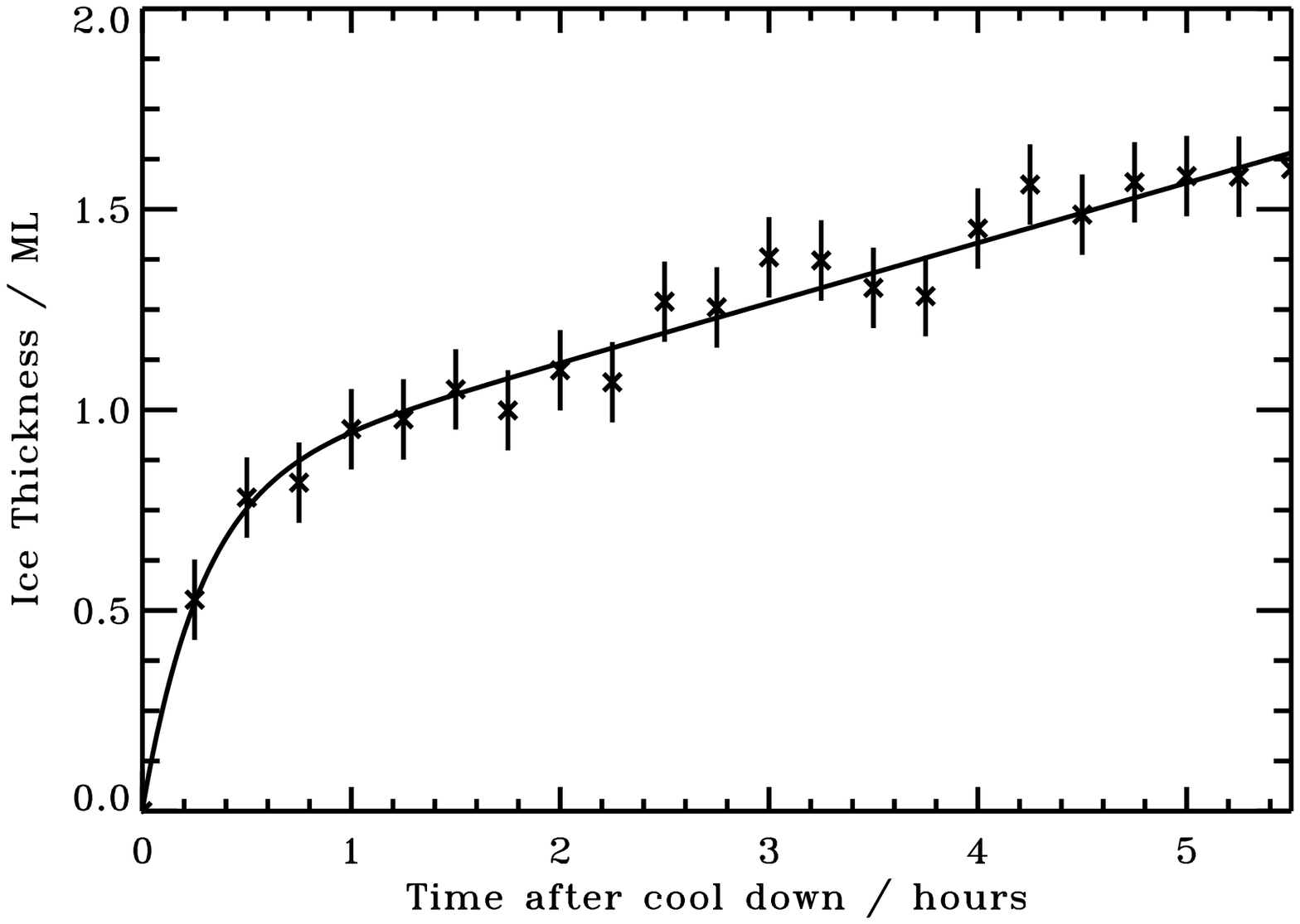}
\caption{The build-up of H$_2$O following cool down to 18 K (without UV irradiation), which is due to the small H$_2$O contamination always present, also under ultra high vacuum conditions. The non-linear behavior is a result of the time required to reach steady-state between H$_2$O freeze-out on the substrate, the desorption of adsorbed H$_2$O from the chamber walls and the H$_2$O pumping.
\label{h2o_buildup}\\[2ex]}
\end{figure}

Figure \ref{temp} also shows that the onset of photodesorption is immediate, i.e. there is no fluence dependence, which is opposite to what was observed by \citet{Westley95}. This difference may be explained by a non-linear H$_2$O freeze-out during the experiment, which is observed in this experiment (Fig. \ref{h2o_buildup}) and probably present in the Westley experiment as well. Even under UHV conditions there is always some H$_2$O present in the vacuum chamber. In our experiment this leads to an ice deposition rate of $\sim$0.1~ML hr$^{-1}$ at equilibrium, but up to 1 ML hr$^{-1}$ is deposited during the first hour after cool down, which is of the same order as the photodesorption rate presented in \citet{Westley95}. In their set-up it would not have been possible to separate this increased freeze-out rate from a lower desorption rate at the beginning of each experiment. This problem is circumvented here by using D$_2$O for most experiments and by letting the H$_2$O freeze-out reach equilibrium before starting the experiment. 

\subsubsection{Substrate and ice structure effects}

In Fig. \ref{depend}a one of the two additional points at 30~K represents the yield from a 9.5~ML  D$_2$O ice deposited on top of a H$_2$O ice layer, showing that the substrate has no effect on the desorption yield. At 100~K this cannot be investigated due to mixing of the two layered ices, but D$_2$O experiments with different ice thicknesses indicate that the substrate has no effect at any temperature for ices thicker than 2~ML (Fig. \ref{depend}b). 

\begin{figure}
\plotone{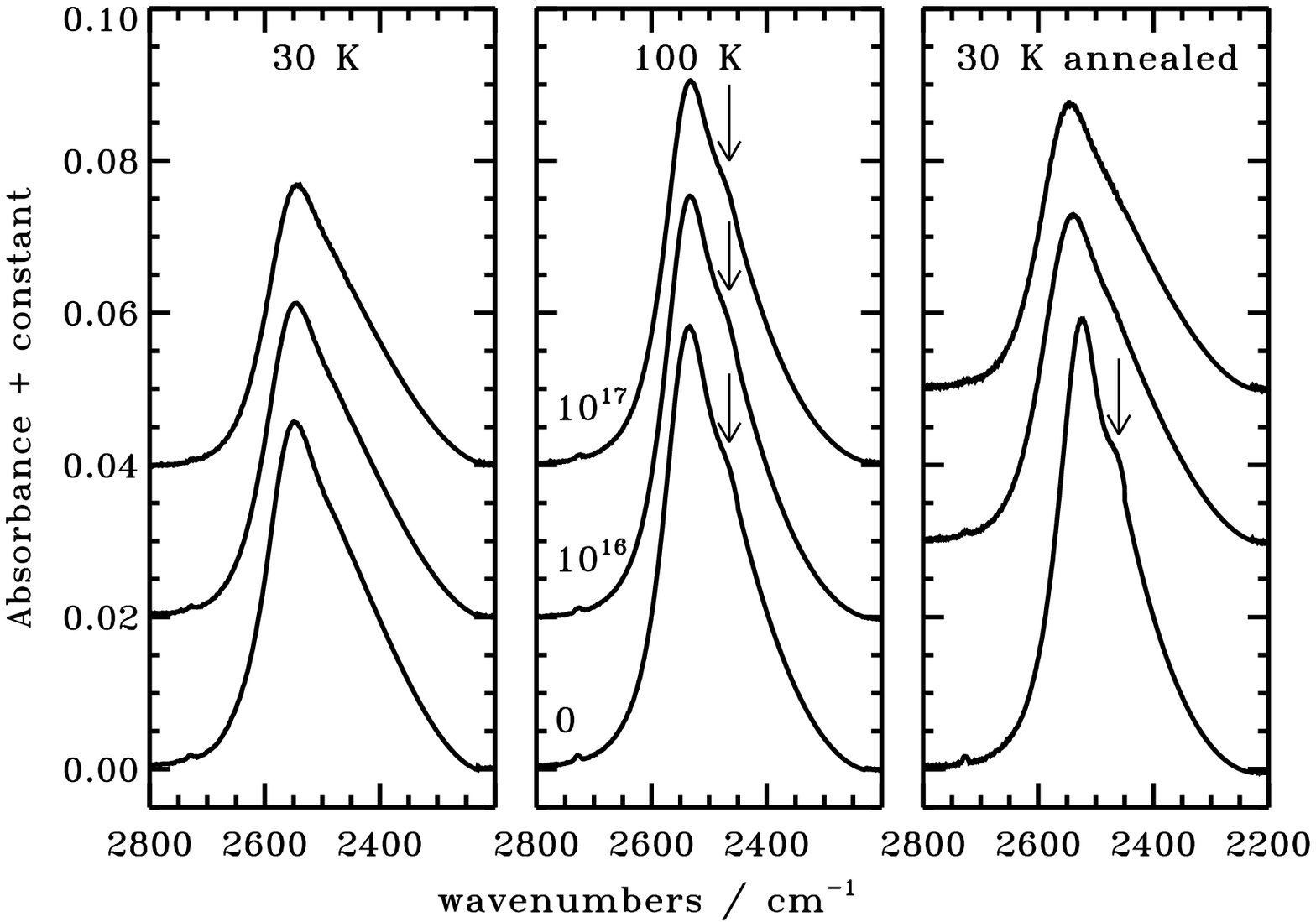}
\caption{Spectra of ices at 30~K (left) and 100~K (middle), and at 30~K, but deposited at 100 K (right), before onset of irradiation and after irradiation by $\sim$10$^{16}$ and 10$^{17}$ photons cm$^{-2}$ as indicated in the middle figure. Note the disappearance of the crystalline feature around 2460 cm$^{-1}$ (marked with arrows) with increasing UV fluence for the annealed ice.
\label{amorph}\\[2ex]}
\end{figure}

The structure independence seen by \citet{Westley95} is also confirmed here for a 30~K ice that is annealed at 100~K, until the spectra display a low frequency shoulder typical for crystalline ice \citep{Hagen81}, and is subsequently cooled down  (the second additional point at 30~K in Fig. \ref{depend}a). Figure \ref{amorph} shows that this can be explained by the fact that the annealed ice returns to an amorphous state upon irradiation with less than 10$^{16}$ UV photons. At 100~K the irradiation does not yield amorphous ice, probably because the temperature is high enough for displaced molecules to diffuse back into a crystalline structure.

\subsubsection{Ice thickness} 

At 100 K, where photodesorption completely dominates over bulk photolysis, there is no photodesorption yield dependence on ice thickness for D$_2$O ices between 8 and 28~ML (Fig. \ref{depend}b) suggesting that the H$_2$O ice photodesorption is only important in the top few layers. At 30~K the photodesorption yield is constant down to at least 3~ML. At both temperatures the thickness dependence is fitted by a function of the type 

\begin{equation}
Y(T,x)=Y_{\rm pd}(T,x>8)(1-e^{-x/l(T)})
\label{thick}
\end{equation}

\noindent where $Y(T,x)$ is the thickness and temperature dependent photodesorption yield, $Y_{\rm pd}(T,x>8)$ the yield for thick ices at a certain temperature, $x$ the ice thickness and $l(T)$ is an ice diffusion parameter, whose origin is discussed further below. The IDL routine $mpfit$ is used to fit the data with the results: $l(100~K) =3.0\pm1.0 \: \rm ML$ and $l(30~K) < 2.7 \: \rm ML$, with the best fit of $l(30~K) =1.3 \: \rm ML$ shown in Fig. \ref{depend}b. Thus photodesorption occurs deeper in the ice at 100~K compared to at 30~K by at least a factor of two. Extrapolating this to lower temperatures  gives

\begin{equation}
l(T) \sim 0.6+0.024\times T
\label{diff}
\end{equation}

\subsubsection{Lamp flux} 

The independence of the photodesorption yield on lamp flux found by \citet{Westley95} for higher fluxes ($1-5\times10^{14}$ photons cm$^{-2}$ s$^{-1}$) is also seen in this study between $1.1-5.0\times10^{13}$ photons cm$^{-2}$ s$^{-1}$ for different temperatures and is shown for 100 K ices in Fig. \ref{depend}c.

\subsubsection{H$_2$O versus D$_2$O} 

Within the experimental uncertainties, there is no difference between the H$_2$O and D$_2$O photodesorption yields at 18 or 100~K (Fig. \ref{depend}d). It should be noted that at 18 K the photodesorption yield of H$_2$O is highly uncertain, because the H$_2$O freeze-out rate dominates over the photodesorption rate during the experiment. Nevertheless, these experiments support the direct applicability of above results, on D$_2$O ices, to H$_2$O ices. 

\section{Discussion}

\subsection{H$_2$O photodesorption mechanism}

The UV photodesorption mechanism of H$_2$O ice does not depend on flux or substrate and is hence most likely due to direct absorption of UV light by H$_2$O molecules, resulting in dissociation of the molecule into fragments with excess energy. The experiments show that once dissociated one of four different outcomes ensues: the dissociated fragments (i) photodesorb directly, (ii) recombine and photodesorb or kick out a surface H$_2$O molecule, (iii) freeze out in the ice or (iv) recombine and freeze out in the ice. These are the same outcomes described in \citet{Andersson06} when performing molecular dynamics simulations on H$_2$O photodesorption. At high temperatures (100~K) the dissociated OH or O fragments are also mobile enough to recombine to e.g. O$_2$. 

The importance of the different pathways depends both on the temperature and where in the ice the molecule is dissociated. With increasing temperature the mobility of molecules, atoms and radicals is expected to increase. This, among other things, increases the reaction probability of OH and H. Assuming the same dissociation yield at all temperatures, the amount of radicals in the ice should then decrease with temperature since the increased mobility with temperature increases the recombination rate. This agrees well with the observed decrease in the steady state photolysis yield between 18 to 100 K (Fig. \ref{temp}). The increased recombination rate also explains the increasing H$_2$O/OH gas phase ratio with temperature during photodesorption. Finally, the increased mobility of OH may also account for the observed photodesorbed O$_2$ at 100 K and its absence at temperatures below 30 K. Although it cannot be excluded that the O$_2$ forms also at lower temperatures and is thermally desorbed following formation above 30 K \citep{Acharyya07}.

The increased mobility with temperature is also reflected in the ice thickness experiments, where the photodesorption occurs down to greater depths in the ice at 100 K compared to 30 K. The factor of two or larger penetration depth into the ice at 100 K compared to 30 K (Fig. \ref{depend}b) agrees well with the increase of the photodesorption yield between 30 and 100 K (Fig. \ref{depend}a). The increased photodesorption yield with temperature is then most likely due to an increased mobility rather than the overcoming of reaction barriers, as suggested by \citet{Westley95b}.

The simulations of \citet{Andersson06} for an ice at 10 K indicate that photodesorption is only efficient in the top 2--3 layers for cold ices. At larger depths freeze-out of the dissociation products completely dominates. This is in excellent agreement with the results of the thickness dependent experiments at 30 K, where photodesorption is only important in the top 3 ML. The simulation is run at ps time scales, while the experiments cover several hours. The agreement for low temperatures between theory and experiment hence indicates that only short time scale processes matter for determining the photodesorption yield at temperatures below 30 K. At higher temperatures longer time scale processes, like thermal diffusion and desorption, increase in importance.

As mentioned above, the depth at which photodesorption occurs increases with temperature, but still there is a certain ice depth where freeze-out of the recombined H$_2$O is the only outcome. At 100 K the measurements are accurate enough to confirm that the mobility of the molecules following photodissociation and recombination is well described by a mean-free-path type model $c\times(1-e^{\rm-x/l(T)})$, where $x$ is the ice thickness, $l$ the diffusion mean-free-path and $c$ the maximum desorption yield for infinitely thick ice. This is also the case for CO$_2$ photodesorption fragments \citep{Oberg08b} and may hence be a universal feature for molecules that photodesorb following dissociation. This mean-free-path behaviour is best explained with the desorption of the energetic photodissocation product or recombined molecule itself, but does not exclude the other outcome of the molecular simulations, i.e. that H$_2$O molecules also desorb indirectly from a kick of a H atom, which originates from photodissociation of another H$_2$O molecule.

\subsection{Comparison with previous experiments}

The results here agree on several important points with those of \citet{Westley95,Westley95b}. The maximum total photodesorption yields (H$_2$O $+$ OH) are the same within the reported uncertainties of both experiments. The determined photodesorption yield is likely a robust value, which is not significantly affected by different lamp spectral energy distributions, order of magnitude different UV fluxes and ice thicknesses. The present study also agrees with \citet{Westley95,Westley95b} on the photodesorption yield dependence on temperature and on the identification of the main desorption products -- H$_2$, O$_2$ and H$_2$O -- with the one exception that we also detect OH.  

The apparent fluence dependence in the Westley experiment can be explained with H$_2$O freeze-out during the early stages of the experiment, especially since they mention a large H$_2$O background pressure in their experiment. This is also in agreement with the mass spectrometer measurements of desorbed species shown in their paper (H$_2$ and O$_2$), which do not show any fluence dependent yields. The apparent fluence dependence led \citet{Westley95} to suggest that at low temperatures desorption occurs through reactions between O and OH. They subsequently claimed that while low temperature photodesorption occurs through photochemistry, high temperature photodesorption is a direct process. From the experiments here it is more likely that both low and high temperature photodesorption processes are dominated by direct photodesorption, but at high temperatures there is some additional desorption due to photochemistry of OH and O fragments.

Ion sputtering of ices has been more thoroughly investigated than ice photodesorption and recent experiments by \citet{Fama08} on the temperature dependence of ion sputtering of H$_2$O ice suggest that the desorption mechanism is comparable for photodesorption and ion sputtering following the initial excitation by a photon or an ion. In particular, in both photodesorption and ion sputtering experiments, the desorption seems highly dependent on the formation and subsequent behavior of radicals and molecular products in the ice. More results are however required on e.g. the thickness dependence and the resulting desorption products during ion sputtering to make an actual comparison between the two processes. The absolute ion-sputtering yield of H$_2$O depends on both the ion energy and ionic species, but it is generally a factor of 10$^3$ to 10$^4$ higher than photodesorption yields i.e. close to unity. When evaluating the importance of the two processes in an astrophysical setting, it is important to note that the ion flux in most regions is orders of magnitude lower than the UV flux. 

\subsection{Astrophysical consequences}

This study shows that pure H$_2$O ice photodesorbs directly or indirectly following fast intermediate photochemistry during which the the photodissociated fragments recombine. The mechanism hence does not depend on the photon flux level or on build-up of radicals in the ice. This means that the yield derived in the laboratory can be directly applied to astrophysical environments. Deep into clouds and disks the rate may be considerably reduced due to e.g. CO ice covers. At the edges of clouds and disks, where other ices have not yet formed at large abundances, the rate for pure H$_2$O ice is directly applicable. Here, as a test case, it is applied to the disk surrounding a Herbig Ae/Be star using models developed by \citet{Dullemond01, Dullemond04} to fit the observed 
   spectral energy distributions of these objects. In the model the physical disk model is static and the chemistry is kept very simple, only including H$_2$O freeze-out, thermal and non-thermal desorption, and no gas phase chemical network except for the recombination of OH to form H$_2$O. 

The model star has a mass of 2.5 M$_\odot$, a radius of 2.0 R$_\odot$ and an effective temperature of 10500 K, typical of a Herbig Ae/Be star, and it emits a pure blackbody spectrum. The accompanying disk has a mass of 0.01 M$_\odot$, with an $R^{-1}$ surface density profile, and an outer radius of 300 AU. To avoid a sharp truncation, the surface density decreases as $R^{-12}$ beyond the outer radius. The inner radius is set by a dust evaporation temperature of 1700 K. The radiation field and dust temperature throughout the disk are calculated using the radiative transfer package RADMC \citep{Dullemond04} and the resulting disk is in vertical hydrostatic equilibrium, with a flaring shape. The gas temperature is set equal to the dust temperature.

\begin{figure}
\plotone{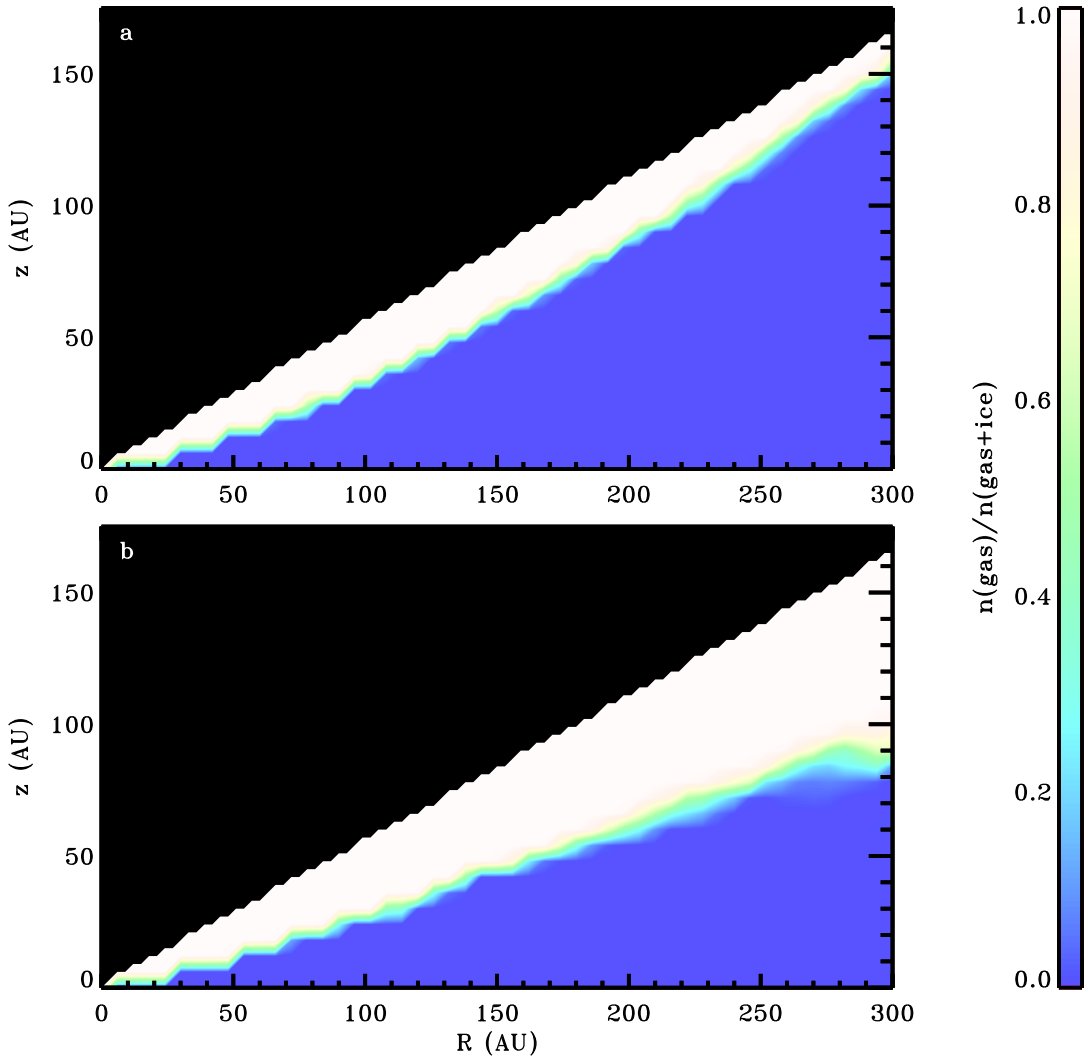}
\caption{Simulation of the distribution of gas phase over total H$_2$O ratio in a circumstellar disk without (a) and with (b) photodesorption. The white, H$_2$O-gas-dominated area extends more than 50 AU deeper into the disk when photodesorption is included, illustrating the large impact of photodesorption on the chemistry in the outer parts of disks.
\label{pd}\\[2ex]}
\end{figure}

Gas phase H$_2$O is initially distributed uniformly throughout the disk at a constant abundance of $1.8\times10^{-4}\times n_{\rm H}$, where $n_{\rm H}$ is the total number of hydrogen nuclei. This is somewhat artificial since H$_2$O forms on grain surfaces, but if the model is run long enough (here to $\sim10^6$ years) the final distribution will not depend on the initial distribution between the gas and grain. The H$_2$O freezes out or adsorbs onto grain surfaces with the rate coefficient $k_{\rm ads}$:

\begin{equation}
k_{\rm ads}=\left(\rm 4.55\times 10^{-18}cm^3 K^{-1/2} s^{-1} \right)n_{\rm H}\sqrt{\frac{T_{\rm g}}{M}}
\label{ads}
\end{equation}
                           
\noindent where $T_{\rm g}$ is the gas temperature and $M$ the molecular weight of water. The numerical factor assumes unit sticking efficiency, a mean grain radius of 0.1 $\mu$m and a grain abundance of $10^{-12}$ with respect to H$_2$ \citep{Charnley01}. Once adsorbed onto the grains, the H$_2$O desorbs thermally with a rate coefficient $k_{\rm thd}$:

\begin{equation}
k_{\rm thd}=\left(\rm 1.26\times 10^{-21}cm^2 \right)A\frac{n_{\rm H}}{n_{\rm s}}e^{-\frac{E_{\rm b}}{ kT_{\rm gr}}}
\label{thd}
\end{equation}

\noindent where $n_{\rm s}$ is the number density of solid water and $T_{\rm gr}$ is the grain temperature. The numerical factor assumes the same grain properties as in Eq. \ref{ads} and 10$^6$ binding sites per grain. The pre-exponential factor, $A$, and the binding energy, $E_{\rm b}/k$, are set to $1\times10^{30}$ cm$^{-2}$ s$^{-1}$ and 5773 K, respectively \citep{Fraser01}. Finally the H$_2$O photodesorption rate coefficient $k_{\rm pd}$ is defined as:

\begin{equation}
k_{\rm pd}=\left(\rm 3.14\times 10^{-14}s^{-1} \right)G_{\rm 0}n_{\rm H}Y_{\rm pd}
\label{pd_eq}
\end{equation}

\noindent where the numerical factor describes the UV photon flux onto a grain surface per unit time for the average interstellar field ($10^8$ photons cm$^{-2}$ s$^{-1} $), $G_{\rm 0}$ is the scaling factor for the UV field that is output by RADMC for each grid point and $Y_{\rm pd}$ is the photodesorption yield. In addition to the external UV field a cosmic ray induced field is approximated by setting a lower limit on $G_{\rm 0}$ of 10$^{-4}$ \citep{Shen04}. The photodesorption results in the release of both OH and H$_2$O. The released OH is quickly rehydrogenated in the model in the gas, however, and hence we let all H$_2$O ices desorb as H$_2$O molecules with the yield:

\begin{eqnarray}
Y_{\rm pd,H_2O} (T_{\rm gr},x)=10^{-3}\left(1.3+0.032\times T_{\rm gr}\right)\left(1-e^{-x/l(T)}\right)
\label{yield}
\end{eqnarray}

\noindent where $x$ is the ice thickness in ML and $l(T)$ the temperature-dependent diffusion length in ML (Eq. \ref{diff}). The model is run for two scenarios - (i) without and (ii) with photodesorption. Each run is 3 Myrs, the typical age of a Herbig disk, which results in steady state gas and grain phase abundances. Figure \ref{pd}a shows the gas phase H$_2$O fraction in the disk without photodesorption as a function of radial and vertical distance from the central protostar. As expected the H$_2$O is completely frozen out, except in the surface layer, when non-thermal desorption is excluded (Fig. \ref{pd}a). When photodesorption is turned on, H$_2$O is kept in the gas phase further in toward the mid-plane (Fig. \ref{pd}b). Without photodesorption 0.6\% of the H$_2$O in the disk is in the gas phase at temperatures above the thermal desorption temperature of 100 K and 0.002\% is in the gas phase below 100 K. With photodesorption included in the model, 0.6\% of the H$_2$O is still present as warm gas, but now 2\% of the total H$_2$O is in the gas phase at temperatures below 100 K. The total column density of warm $T>$100 K H$_2$O gas, averaged over the entire disk, is hence the same in both cases ($1.4\times10^{17}$ cm$^{-2}$). In contrast the amount of cold H$_2$O gas averaged over the entire disk increases from $5.0\times10^{14}$ to $4.5\times10^{17}$ cm$^{-2}$ when photodesorption is turned on. This means that a gas phase chemistry involving OH or H$_2$O is possible deep toward the disk midplane also in the outer disk. For comparison we also run our model using a constant surface photodesorption yield of 10$^{-3}$, which has often been used in the literature previously. For this particular model the total column density changes with less than a factor of two compared to using the derived yield equation from this study. The spatial distribution of gas phase H$_2$O is however different using the different yields due to the fact that using the constant yield overestimates the desorption rate in the surface region and underestimates it deeper into the disk.

This is a generic disk model commonly used to model disks around intermediate mass stars. To model an actually observed object would require a more detailed model that takes into account observed constraints on the disk structure. In addition, the calculated gas phase abundances may change somewhat when chemistry is taken into account. The general trend is however that photodesorption increases the amount of cold gas phase H$_2$O by orders of magnitude. This is also the result of a recent PDR model showing that the inclusion of chemistry, while important for more accurate predictions, will not reduce the predicted column density of gas phase H$_2$O dramatically (Hollenbach et al., ApJ, in press). Other effects such as grain growth may increase the photodesorption rate, but without a full chemical network it is unclear how much of this increase will be off-set by photodissociation of the desorbed H$_2$O. Probably only observations of cold H$_2$O on the scale of protoplanetary disks will yield an answer. The beam of the imminent {\it Herschel Space Observatory} is of the same order of magnitude as this modelled disk and hence these results show that large amounts of cold H$_2$O will be observable. For exactly this purpose -- to observe cold H$_2$O gas -- the WISH program was approved as a {\it Herschel} key program. The amount of gas phase H$_2$O in disks due to non-thermal desorption may hence be answered very soon indeed.

\section{Conclusions}

   \begin{enumerate}
      \item The total D$_2$O  and H$_2$O photodesorption yields are indistinguishable within the experimental uncertainties and are empirically described by $Y_{\rm pd}(T,x)=Y_{\rm pd}(T,x>8)(1-e^{-x/l(T)})$ where $Y_{\rm pd}(T,x)$ is the thickness and temperature dependent photodesorption yield, $x$ the ice thickness in monolayers and $l(T)$ an ice diffusion parameter that varies between 1.3 ML at 30~K and 3.0 ML at 100~K. 	
      
        \item For thick ices ($>$8 ML), the yield depends linearly on temperature such that $Y_{\rm pd}(T,x>8) = 10^{-3}\left(1.3+0.032\times T\right)$ photon$^{-1}$. The yields agrees, within the reported 60\% uncertainty, with a previous experiment \citep{Westley95}.
      
      \item The nature of the desorbed species is temperature dependent, with equal amounts of OH and H$_2$O detected at low temperatures. At higher temperatures the H$_2$O:OH fraction is $\sim$ 2:1 and in addition about a fifth of the ice photodesorbs as heavier fragments like O$_2$. The fraction of the total photodesorption yield that results in H$_2$O molecules desorbing is described by $f_{\rm H_2O} =  (0.42\pm0.07)+(0.002\pm0.001)\times T$.
        
      \item We find no yield dependence on photon flux or fluence. The fluence independence is in contrast with a previous experiment \citep{Westley95}.
      
      \item We also find no dependence on the ice structure i.e. whether the D$_2$O ice is amorphous or crystalline. This is consistent with spectroscopic evidence of fast destruction of crystalline ice into an amorphous state following UV irradiation.
      
      \item The photodesorption yield and dependencies found here are consistent with previous theoretical predictions of H$_2$O photodesorption, where the photodesorption is limited to the top few layers of the ice \citep{Andersson06}. In addition we see that the photodesorption yield increases with ice temperature because of the increased mobility of the photolysis fragments, allowing desorption from deeper within the ice.
      
      \item Applying the experimental yield to a Herbig Ae/Be star+disk model we calculate that the predicted amount of cold ($<100$ K) gas phase H$_2$O, averaged over the entire disk, increases with orders of magnitude due to photodesorption.
   \end{enumerate}

\bibliographystyle{aa}
%\bibliography{PhotodesRefs.bib}

\acknowledgments

The authors wish to thank Stefan Andersson and Herma Cuppen for stimulating discussions. Funding is provided by NOVA, the Netherlands Research School for Astronomy, a grant from the European Early Stage Training Network ('EARA' MEST-CT-2004-504604) and a Netherlands Organisation for Scientific Research (NWO) Spinoza grant.

\end{document}